\documentclass{pasj00}

\begin{document}
\draft
\newcommand{\gsim}{\raisebox{-.4ex}{$\stackrel{>}{\scriptstyle \sim}$}}
\newcommand{\lsim}{\raisebox{-.4ex}{$\stackrel{<}{\scriptstyle \sim}$}}
\newcommand{\psim}{\raisebox{-.4ex}{$\stackrel{\propto}{\scriptstyle \sim}$}}
\newcommand{\kms}{\mbox{km~s$^{-1}$}}
\newcommand{\s}{\mbox{$''$}}
\newcommand{\mloss}{\mbox{$\dot{M}$}}
\newcommand{\my}{\mbox{$M_{\odot}$~yr$^{-1}$}}
\newcommand{\ls}{\mbox{$L_{\odot}$}}
\newcommand{\ms}{\mbox{$M_{\odot}$}}
\newcommand{\mm}{\mbox{$\mu$m}}
\def\arcdeg{\hbox{$^\circ$}}
\newcommand{\secp}{\mbox{\rlap{.}$''$}}%

\SetRunningHead{Rioja, Dodson, Kamohara, Colomer, Bujarrabal, Kobayashi}
{Astrometric observations of SiO masers in R\,LMi using VERA}
\Received{2008/xx/xx}
\Accepted{2008/xx/xx}

\title{Relative astrometry of the J=1$\rightarrow$0, $v$=1 and $v$=2
  SiO masers towards R Leonis Minoris using VERA}

 \author{Mar\'{\i}a J. \textsc{Rioja}\altaffilmark{1,2}
   Richard \textsc{Dodson}\altaffilmark{1,2}
   Ryuichi \textsc{Kamohara}\altaffilmark{3}\\
   Francisco \textsc{Colomer}\altaffilmark{2}
   Valent\'{\i}n \textsc{Bujarrabal}\altaffilmark{2}
   and
   Hideyuki \textsc{Kobayashi}\altaffilmark{3}}
 \altaffiltext{1}{School of Physics, University of Western Australia, Australia}
 \altaffiltext{2}{OAN, Alcal\'a de Henares, Madrid, Spain}
 \altaffiltext{3}{VERA Project Office, NAOJ, Mitaka, Tokyo, Japan}
 \email{mj.rioja@oan.es}

\KeyWords{Stars: AGB and post-AGB -- Stars: circumstellar matter -- 
          Stars: late-type -- Masers (SiO) -- VERA} 

\maketitle

\begin{abstract}

  Oxygen-rich Asymptotic Giant Branch (AGB) stars are intense emitters
  of SiO and H$_2$O maser lines at 43 (J=1$\rightarrow$0, $v$=1 and
  2) and 22 GHz, respectively. VLBI observations of the maser emission
  provides a unique tool to sample the innermost layers of the
  circumstellar envelopes in AGB stars.  Nevertheless, the
  difficulties in achieving astrometrically aligned $v$=1 and
  $v$=2 SiO maser maps have traditionally prevented a unique
  interpretation of the observations in terms of physical underlying
  conditions, which depend on the nature of the SiO
  pumping mechanism.
  We have carried out observations of the SiO and H$_2$O maser
  emission towards R\,LMi, using the astrometric capabilities of
  VERA. Due to the too-weak emission of the reference calibrator we
  had to develop a special method to accurately relate the coordinates
  for both transitions. We present relative astrometrically aligned $v$=1 and
  $v$=2 J=1$\rightarrow$0 SiO maser maps, at multiple epochs, and
  discuss the astrophysical results.  The incorporation of astrometric
  information into the maps of SiO masers challenges the weak points
  in the current theoretical models, which will need further
  refinements to address the observations results.

\end{abstract}

\section{Introduction}

O-rich AGB stars are intense emitters of SiO and H$_2$O maser lines,
particularly at 43 (SiO lines, $J$=1$\rightarrow$0, $v$=1 and 2) and
22 GHz (H$_2$O, $J_{K_-,K_+}$=6$_{1,6}$--5$_{2,3}$). SiO masers
require a very high excitation and appear at a distance of a few
stellar radii, $\sim$10$^{14}$ cm, occupying a more or less circular
structure. H$_2$O emission is found further from the star,
$\sim$10$^{15}$ cm, with less well defined structures. A combined study
of both masers should then produce a very accurate description of the
structure and kinetics of these inner layers. In such inner
circumstellar shells, from which the whole circumstellar envelope will
be formed, the dust grains are still growing and the gas has not yet
attained its final expansion velocity, since expansion is supposed to
be powered by radiation pressure onto grains. The dynamics in the SiO
emitting region is dominated by pulsation, which propagates from the
photosphere via shocks, and by the first stages of outwards
acceleration. Their study is therefore basic in understanding the mass
loss process and the envelope formation in AGB stars.

The relative positions of the spots of both $v$=1 and $v$=2 SiO
transitions are an important result in order to elucidate the nature of
the pumping mechanisms (Humphreys \etal\ 2002, Bujarrabal
1994). Existent maps suggest that the spots of these lines are
syste\-ma\-ti\-ca\-lly separated by a few milliarcsecond (mas),
(Desmurs \etal\ 2000, Yi \etal\ 2005, Boboltz \& Wittkowski 2005). In
the former, the authors have explained why we believe this favors
radiative mechanisms, but we note that Yi \etal\ found similar
offsets, but also many common features, which they argued supported the
collisional model.
A few attempts to study theoretically the observed spatial
distribution of the different SiO masers have been published, see the
short report on collisional pumping predictions for the relative
distribution of $v$=1,2 $J$=1$\rightarrow$0 and $v$=1
$J$=2$\rightarrow$1 masers by Humphreys \etal\ (2000), and the
predictions for the same parameters from radiative models by
Soria-Ruiz \etal\ (2004), including in this case the effects of line
overlap.  However, existing models are not sophisticated enough to
undertake a sensible explanation of present high-quality maps of SiO
maser emission, as we will discuss in Sect.\ 4.

Our understanding of these problems has greatly changed from the first
empirical comparisons between conventional VLBI observations, which do
not provide us with absolute coordinates. Accordingly, the location
and shape of the maser regions, one with respect to the other and with
respect to the star's position [very accurately measured for many AGB
stars, with Hipparcos satellite (Perryman \etal\ 1997) with a typical
precision of 1\,mas], are still not well known. The studies of SiO and
H$_2$O maser spot distributions are therefore severely hampered by the
lack of accurate information on the absolute coordinates in the maps
of the different lines. 

The dual-beam interferometer VERA is very well suited to produce VLBI
maps with accurate absolute astrometry, thanks to the dual-beam design
that enables simultaneous observation of the maser line and a
reference continuum source (in general, a quasar), and its inherent
instrumental phase stability.  This approach relies on the
detectability of the reference source, which in practice can be
problematic.  In the observations presented in this paper absolute
astrometry with respect to a celestial reference frame was not
possible due to the weakness of the simultaneously observed reference
continuum source.  We present here multi-epoch VERA observations of
SiO and H$_2$O maser emission in the well known Mira-variable R\,LMi,
and the relative alignement of $v$=1 and $v$=2 J=1$\rightarrow$0 SiO
maser emission derived from an astrometric analysis.


We describe our data reduction procedure to partially recover
astrometric information, even when there are no frequent observations
of a reference source. This method is applied to our observations of
R\,LMi, and was shown to be useful to measure the relative positioning
of both SiO maser lines. Results from the analysis of four different
epochs of observations are presented.

R\,LMi is an O-rich Mira-type variable, with a pulsation period of about
373 days and a spectral type ranging between M6.5, at the optical
maximum, and M9.0, at the minimum. Its distance is assumed to be $\sim$
350 pc from the well-known period-luminosity relation, the Hipparcos
parallax measurement being poor (see Whitelock \etal , 2000). R\,LMi is
a well known emitter in SiO masers, it was in particular
accurately monitored by Pardo \etal\ (2004), who found the periodic
variation in-phase with the IR cycle typical of Mira-type
stars. However, accurate VLBI maps of SiO masers in R\,LMi had not been
reported up to date.

The contents of the paper are organized as follows.  The observational
setup is described in Sect.\ 2. In Sect.\ 3 we describe the special
data reduction strategy used to preserve the relative astrometry
between the two SiO maser transitions. In Sect.\ 4 we present
the resulting maps and discuss the astrophysical results.

\section{Observations}

\begin{figure}
  \begin{center}
 \FigureFile(80mm,80mm){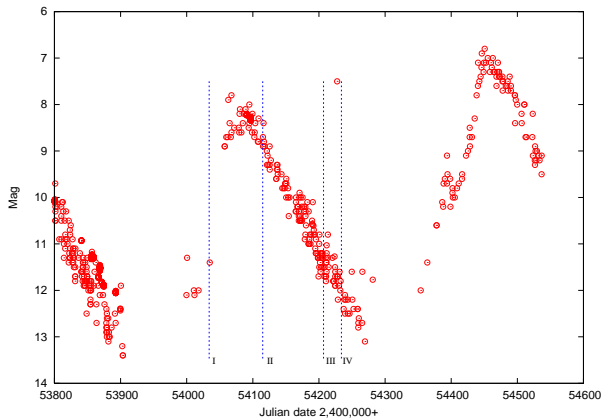}
  \end{center}
  \caption{
Stellar optical light curve for R\,LMi (courtesy of AAVSO).
The vertical lines mark the epochs of the SiO maser observations, with VERA,
presented in this paper.}
\label{fig:aavso}
\end{figure}

We carried out a series of VERA observations at 5 epochs of the SiO
maser emission ($v$=1 and $v$=2 J=1$\rightarrow$0), at 43 GHz, and
3 epochs of observations of H$_2$O maser emission, at 22 GHz, towards
R\,LMi, between October 2006 and May 2007. Of these, four epochs of SiO and
one epoch of H$_2$O maser observations have been completely analyzed, 
and the results are presented in this paper.  Table 1 lists the epoch
and duration of the observations 
along with the s\-te\-llar phases, as derived from the optical light curve 
shown in Fig.~\ref{fig:aavso}.

At all epochs the observations were made in the dual-beam mode, with
the 4 VERA antennas (Mizusawa, Iriki, Ishigaki-jima, and Ogasawara)
observing simultaneously R\,LMi and a nearby continuum reference source
(J0945+3534 or J0952+3512) for $\sim$8-9 hours.  Every hour, a $\sim$5 min scan
on a bright continuum fringe-finder calibrator source, 4C39.25
(J0927+3902), was included.

The observations were performed with the DIR2000 recording system at a
data rate of 1 Gbps.  Each antenna recorded an aggregate of 256 MHz
bandwidth, using 2-bit Nyquist sampling, subdivided into a total of 16
IF channels for both beams.
During the observations at 43 GHz, two 16-MHz channels were allocated
to the observations of the two SiO maser transitions with one beam
(beam A).  The IF channels were tuned to match the Doppler-shifted
frequencies of the $v$=2, J=1$\rightarrow$0 (rest frequency
42.820584 GHz) and the $v$=1, J=1$\rightarrow$0 (rest frequency
43.122027 GHz) line transitions.  The remaining fourteen 16-MHz IF
channels were devoted to the simultaneous observations of the
continuum source on the other beam (beam B), evenly distributed over
the $\sim$300 MHz frequency gap between the two maser transitions.
During the 22 GHz observations, only one 16-MHz IF channel (centered on
22.3512\,GHz) was allocated to the observations of the H$_2$O maser
emission (beam A). The other fifteen IF channels were allocated to the
simultaneous observations of the continuum source with the other beam
(beam B).

The correlation was done at the Mitaka FX correlator (Chikada \etal\ 1991)
with a spectral resolution of 512 channels per IF for beam A, and 64 channels
per IF for beam B. At a single epoch, output data sets were
generated for each beam, consisting of the visibility functions averaged
to 1 second, with samples every 31.25 kHz in frequency, yielding 
velocity resolutions of 0.22\,\kms and 0.4\,\kms 
for the line observations at 43 and 22 GHz, respectively.

\begin{table}
\caption{Details on the SiO and H$_2$O maser observations towards R\,LMi
with VERA.}\label{tab:o}
\begin{center}
\begin{tabular}{lcccr}
Epoch &Year/DOY &Duration (UT) & Stellar Phase&\\
\hline
\hline
\multicolumn{5}{c}{SiO}\\
I   & 2006/298 & 18:30-02:30 & $\phi = 0.84$&\\
II  & 2007/014 & 12:30-21:30 & $\phi = 0.07$&\\
III & 2007/106 & 06:30-15:30 & $\phi = 0.32$&\\
IV  & 2007/133 & 05:00-14:00 & $\phi = 0.40$&\\
\multicolumn{5}{c}{H$_2$O}\\
  & 2007/132 & 05:00-14:00 & $\phi = 0.40$&\\
\hline
\hline
\end{tabular}
\end{center}
\end{table}

\section {Data reduction}

The ``conventional'' phase referencing analysis strategy with VERA
consists of transferring phase/delay/rate solutions, from the data
reduction of observations of the reference source on one beam to the
simultaneous observations of the target source on the other beam.
The calibrated visibilities of the target source are Fourier
transformed without further calibration to yield synthesis images,
which preserve the astrometric information with respect to an external
reference.

Unfortunately, the continuum reference source in our observations
was too weak to be detected using self-calibration techniques, at either
of the frequencies, and hence prevented us from following the
``conventional'' analysis route.  Some authors have performed the
``inverse'' analysis, and used the data from the line source to
enhance the detection of a weak continuum reference source (Choi, Kamohara,
priv.\ comm.); this approach also failed in our observations.

The lack of detection of the continuum reference source prevents us
from achieving our initial goal, that is measuring absolute positions
of all the maser maps, i.e. the H$_2$O (22\,GHz) and two SiO (43.1\,GHz
and 42.8\,GHz) maser transitions, with respect to an external
reference, and with respect to the star.
Instead, we have followed a different analysis strategy that allows
the measurement of the alignment between multiple transitions
simultaneously observed, even in cases when the reference source is
not detected, using the interleaved observations ($\sim$every hour) of
the primary calibrator source, 4C39.25, $5.9^o$ away.  Our proposed
analysis route preserves the relative positions between the two SiO
maser maps at each single epoch, but we can not recover the astrometry
with respect to the external continuum source on beam B, thus
preventing comparison between different epochs and with respect to the
H$_2$O maser maps.
 
We used the NRAO AIPS software package for the data reduction. The
information on measured system temperatures, telescope gains, and
estimated bandpass corrections at the individual antennae were used to
calibrate the raw correlation coefficients of the calibrator and,
along with Doppler corrections, of the line source observations.  Then
we used the AIPS task FRING, which is a global self-calibration
algorithm, to estimate residual antenna-based VLBI phases
and its partial derivatives with respect to frequency (hereafter 
group delay, $\tau$), and time (hereafter rate), on the calibrator data
set. These terms result from unaccounted contributions from the
atmospheric propagation and from errors in the array geometry during
the data processing.

\begin{figure}
  \begin{center}
    \FigureFile(80mm,80mm){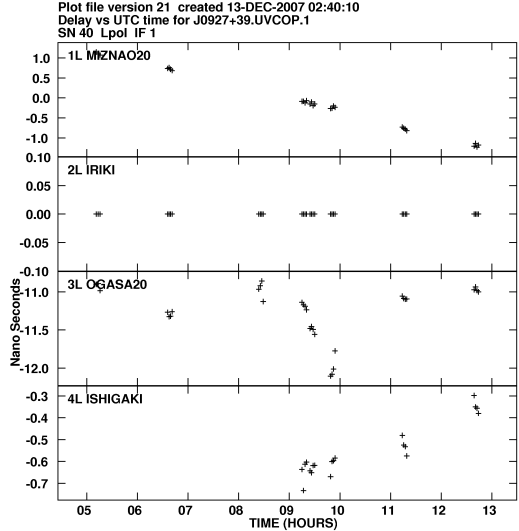}
  \end{center}
   \caption{Temporal variation of residual VERA group delays 
estimated from the analysis of 4C39.25 observations at 43 GHz, with FRING,
for epoch IV. These introduce different phase ambiguities at each IF.} 
\label{fig:delays}
\end{figure}

An advantage of the digital filters in the VERA DIR2000 recording
system is the absence of instrumental phase term offsets introduced by
the electronics at each IF, which enables the straight-forward use of
all the IFs as an effective bandwidth for the estimation of a more
precise group delay (since $\sigma_{\tau}$ $\propto$
1/bandwidth). For arrays with independent IF electronics 
a further stage of calibration would be required.  Fig.~\ref{fig:delays} 
shows an example
of the slow temporal variation of the estimated group delays for the
calibrator source 4C39.25, at one of the epochs of observations and
for all VERA antennas. These residual group delays
measure the rate of change of the phase with frequency, and hence the phase
difference ($\Delta \phi$) between IFs separated by
a bandwidth $\Delta \nu$ is given by $\Delta \phi = 2 \pi \, \Delta
\nu\, \tau (t)$.
The stability of the system allows us to use observations of a
calibrator made every hour or so -- much less frequently than Yi \etal
. The scarcity of calibrator scans requires us to take great care with the
interpolation of the phase from the group delay solutions.

In the analysis of the R\,LMi data set we used the interpolated values
of the calibrator group delays to the times of the line observations to
remove the ``instrumental'' long term phase difference between the two
IFs, in order to preserve the relative coordinates between $v$=1 and
$v$=2 masers. As the group delays vary with time, the relative phase
differences between IFs also change, and need to be tracked to prevent
loss of coherence. Furthermore we found that the calibrator phases,
which are changing essentially randomly between the hourly
scans, could disrupt the phase connection across the 300-MHz spanned bandwidth
by introducing different number of phase ambiguities at each IF
and reduce the success
rate for accurate phase tracking.  Zeroing the calibrator phase
entries is not a solution, since they also carry information on the
relative phase difference between the IFs, as derived from the group delay
estimates.  In the {\it modus operandi} of AIPS, the group
delays are stored as an IF-based delay entry, common for each IF,
which represents the rate of change, and a phase entry for each IF
derived from that delay and its difference to the reference
frequency. Then, the calibrator phases at the reference frequency
are added to those (delay related) phase entries.

\begin{figure}
  \begin{center}
    \FigureFile(80mm,200mm){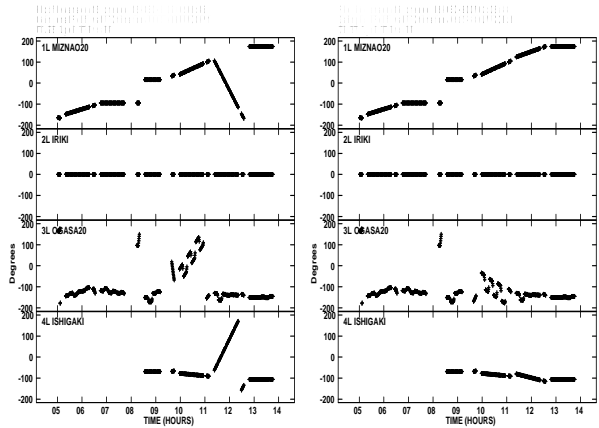}
  \end{center}
  \caption{Differences between antenna-based phase solutions for IFs 
    separated by a frequency gap of $\sim 300$\,MHz,
    at epoch IV.  The phase entries have been estimated with FRING at
    the times of the $\sim$\,hourly continuum calibrator scans, on
    beam B, and interpolated to the times of R\,LMi observations (on
    beam A).  {\it Left} and {\it Right} plots show the phase
    differences before and after removing the phases from the
    calibrator, respectively, whilst retaining the inter-IF terms 
    derived from the group delay (see text).  When the calibrator
    phase terms are not removed the expected, smooth phase
    connection between calibrator scans is occasionally broken.
    At the analysis, equally good results can be obtained using our scripts, 
    or by subtracting out the phases from the reference IF.}
\label{fig:zero}
\end{figure}

To solve the problem with introducing extra phase ambiguities across
the 300-MHz, we exported the solution table produced by FRING on the
calibrator data set, and zeroed the rate and the phase entries for the
selected reference frequency (chosen at 43\,GHz); also, we
recalculated the phase entries for the rest of IFs from its difference
with the reference frequency and the estimated delay. This ensured
that the phases deduced from the delays could be transferred across
the $\sim$300\,MHz between the two observed SiO maser transitions.  If
this was not done, the phase connection would occasionally breakdown
(see Fig.~\ref{fig:zero}).  This problem affected to about 20\% of the
observing time in our data sets.  These extra steps ensured
unambiguous phase connection across the total duration of the
observations and preserved the alignment between the SiO maser
transitions.

Subsequently we have realized that part of the handling of the SN
table can be done using the AIPS task SNCOR, with mode REFP, on a
central IF. This approach has the advantage of not having to work
outside the AIPS environment. Nevertheless we are happy to provide our
perl script on request, as it also will phase center on an arbitrary
frequency (useful when the reference IFs for beam A and beam B are not
at the same frequency), and expand or contract the SN table for the
different numbers of IFs allocated to beams A and B.

Then, we applied spectral line self calibration and imaging techniques
on a compact and bright reference maser spot in the $v$=2 dataset, to
track the short term atmospheric fluctuations and get a map of the
feature.
At each epoch, the resulting antenna-based phase and rate solutions from 
a reference channel
in the $v=$2 IF were used to calibrate all other $v=$2 spectral channels
(the usual case) and also all $v=$1 channels in the other IF without separate
$v=$1 reference channel calibration.
%
The phase-referenced visibilities (of all velocity channels) were Fourier
transformed using the AIPS task IMAGR, without further calibration, 
to yield image cubes of $v=$1 and $v$=2 SiO masers that preserve the relative
astrometric registration with respect to the reference spot (in $v$=2). 

This method allows relative astrometric alignment of the two
frequencies, even if the maps have no absolute coordinates, largely
thanks to the instrumental stability of the digital receiving system
in VERA.  The astrophysical results obtained from the maps and its
astrometric alignment are presented in the next section.

This method would be expected to work when the errors introduced by
uncertainties in the correlator model, the temporal interpolation
between calibrator observations, and the extrapolation between
different directions in the sky ($5.9^o$ away) and across the
$\sim$300 MHz spanning bandwidth between both transitions are
reasonably small. We are confident that the interpolation between
calibrator observations worked well because 4C39.25 (J0927+3902) was
also occasionally observed in beam A, and we get a relative
astrometric alignment between the maps of the two IFs, $\sim300$\,MHz
away, of $<0.1\,$mas. 
This is of course on the same source from which we derive the delays,
but at different times and with a different receiver. The
uncertainties related to the spatial extrapolation result from
unmodelled terms related to the atmospheric contribution, and source
position errors.  While the former will degrade the results adding
noise, the latter propagates into the analysis as a systematic shift
in the astrometric measurements between two SiO transitions maps.

There have been attempts to transfer the solutions between the $v$=1
and 2 maser transitions before, e.g. Boboltz \& Wittkowski (2005)
using observations with only two 8-MHz IFs. The concern is that,
because of the narrow bandwidths and the low accuracy of the estimated
delays, the relative phase between the two transitions might not be
preserved.
Yi \etal\ (2005) describe a wideband method, very similar to ours,
which covers the 300-MHz span. Their attempt, however, failed because
of the poor accuracy of the coordinates of the star observed, TX
Cam. This led to a systematic relative error between the $v$=1 and
$v$=2 positions of about $2\,$mas. They, therefore, assumed a common
feature for the two SiO maser transitions and derived the absolute position of
TX-Cam.
If an independent position had been available the same analysis as
ours could have been applied.
In our case we have observed a wide bandwidth to ensure that the
delay errors are small, and both sources have positions measured with
high precision astrometry (Hipparcos for R\,LMi, and VLBI for 4C39.25),
with accuracies better than $1\,$mas. However the uncertainty in the
proper motion of the line source, given that the Hipparcos
observations were made in the early 1990's, results in a position
error of the order of $30\,$mas. We assume this is the dominant source
of errors in our analysis and will use it to determine the final error
bars in our astrometric measurements.  An absolute position error of R
LMi of $50\,$mas in our analysis would propagate into less than
$0.4\,$mas in the relative shift between the transitions (result from
multiplying the star's position error by the phase-referencing
attenuation factor, $\Delta \nu / \nu$).
With all, we give a conservative estimate of the uncertainties in our
measurements of the relative $v$=1 and $v$=2 maser positions, at all
epochs, of $0.4\,$mas.

\section{Results and discussion}

The map resulting from H$_2$O maser observations towards R\,LMi showed
a single compact component which was slightly resolved and had an
integrated flux density of 3\,Jy. The simple structure in the maps
along with the lack of astrometric information, due to the
non-detection of the reference source, invalidate our attempt to align
the H$_2$O and SiO maser emission in R\,LMi. Because of this we have
only fully analysed one epoch of the H$_2$O maser observations, as shown 
in Table 1.

\begin{figure}
  \begin{center}
    \FigureFile(80mm,80mm){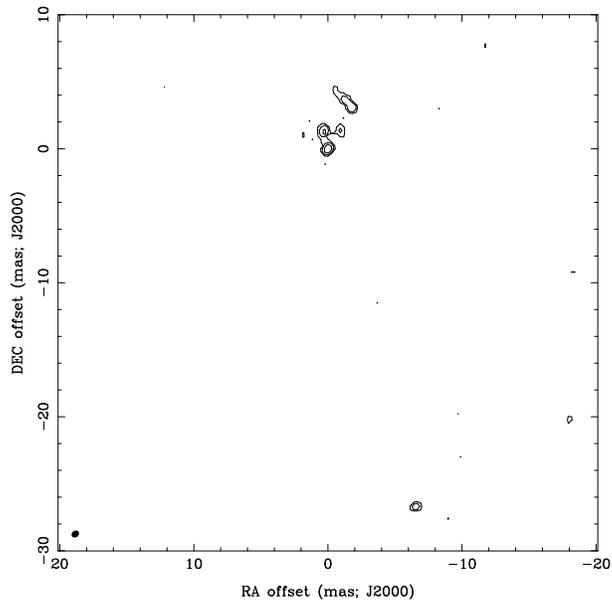}
  \end{center}
  \caption{Contour map of SiO $v$=1 J=1$\rightarrow$0 maser emission
    in R\,LMi in epoch I, averaged over regions with significant
    emission. The contours are at 1, 2, 4 and 6 Jy/beam. The noise
    levels in the cube were 0.2 Jy/bm/channel.}
\label{fig:one}
\end{figure}

\begin{figure}
  \begin{center}
    \FigureFile(80mm,80mm){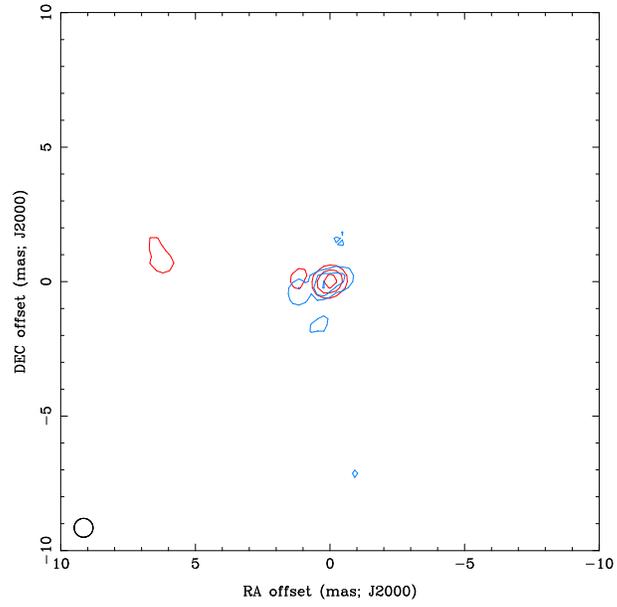}
  \end{center}
  \caption{Relatively astrometrically aligned contour maps of $v$=1
(red; less intense lines) and $v$=2 (blue) J=1$\rightarrow$0 SiO maser
emission towards R\,LMi, with respect to the position of the strongest
channel in $v$=2, at epoch II.  The contours are at 3, 6 and 10
Jy/beam. The noise levels in the cubes were 0.2 Jy/bm/channel.}
\label{fig:two}
\end{figure}

\begin{figure}
  \begin{center}
    \FigureFile(140mm,140mm){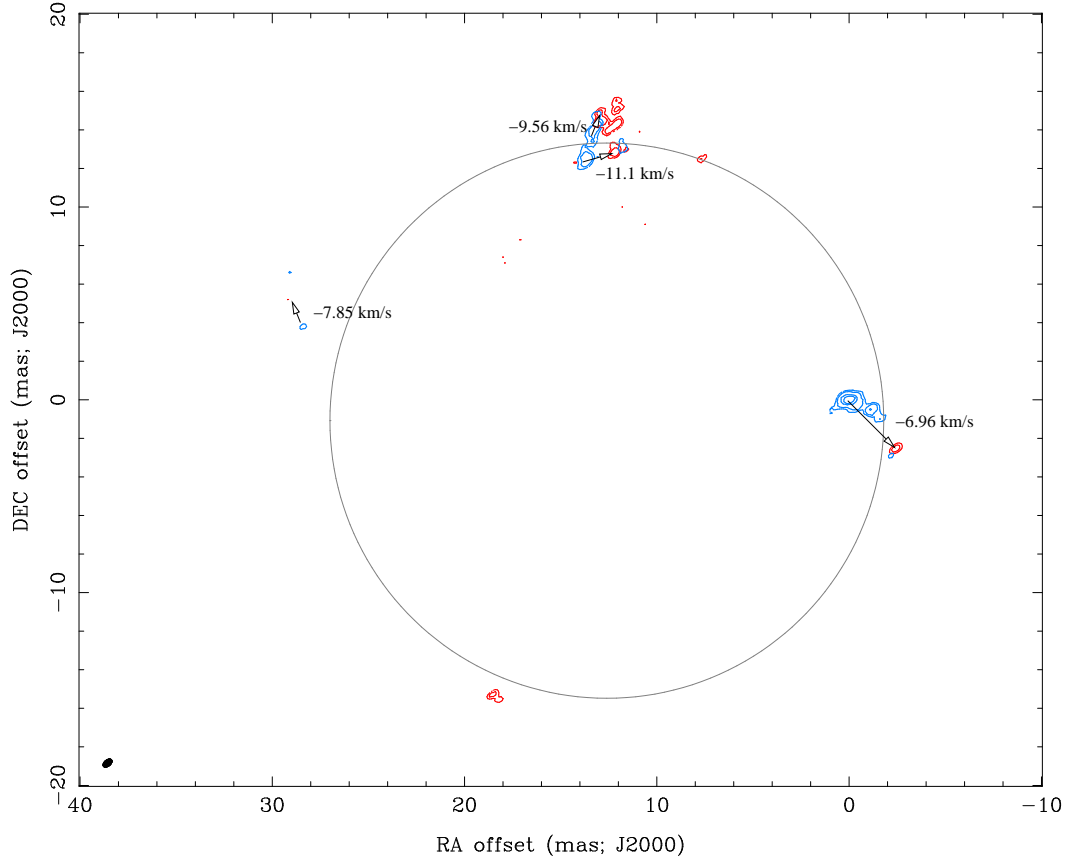}
  \end{center}
  \caption{Relatively astrometrically aligned contour maps of $v$=1
    and $v$=2 J=1$\rightarrow$0 SiO maser emission towards R\,LMi,
    with respect to the position of the strongest channel in $v$=2, at
    epoch III. We used the same colors as in previous figure. The
    contours are 2,\,3,\,6,\,8 and 10 Jy/beam. The noise levels in the
    cubes were 0.2 Jy/bm/channel. The circle approximates the assumed
    ring of emission around the central star; it is not a fitted
    solution, and serves for visual comparison of the spatial
    distribution of spots at different epochs. The arrows indicate the
    relative position between nearby spots of $v$=2 and $v$=1 masers
    with common velocities.  The size of the arrows head corresponds
    to our conservative estimate of the errors in the astrometric
    alignment between the $v$=1 and 2 maser maps (see text). }
\label{fig:three}
\end{figure}

\begin{figure}
  \begin{center}
    \FigureFile(60mm,60mm){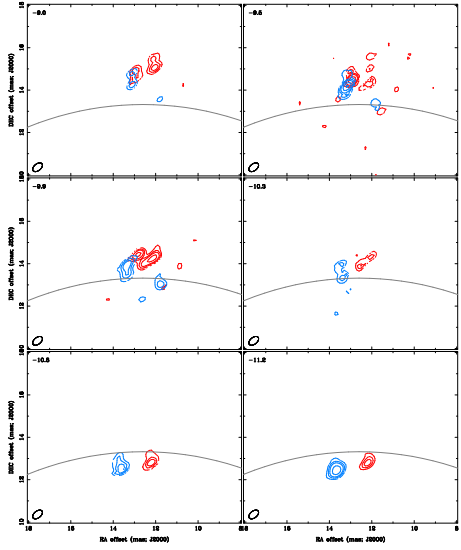}
  \end{center}
  \caption{Detail of the epoch III map shown in Figure 6 (Northern
    cluster), with the channels averaged to 4.4\,\kms. The
    colors are as before. One can see that, whilst there is one
    cluster with common velocity and position, none of the others
    overlap. Contours are at 2, 3, 4, and 6 Jy/beam.}
\label{fig:last}
\end{figure}

\begin{figure}
  \begin{center}
    \FigureFile(80mm,80mm){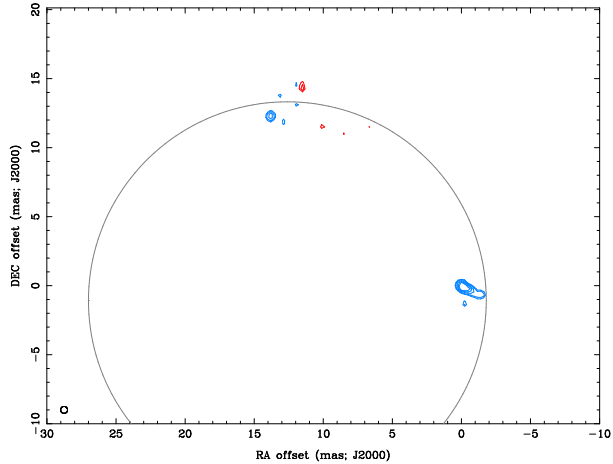}
  \end{center}
  \caption{Relatively astrometrically aligned contour maps of $v$=1
    and $v$=2 J=1$\rightarrow$0 SiO maser emission towards R\,LMi,
    with respect to the strongest channel in $v$=2, at epoch IV. The
    colors are as in previous figures. The contours, for $v$=2, are at
    3, 4, and 6 Jy/beam and, for $v$=1, are 3, 4, 6, 8, 10
    Jy/beam. The circle aims to be of help for a visual comparison of
    the spots distribution between epochs (see text). The noise levels
    in the cubes were 0.3 Jy/bm/channel.}
\label{fig:five}
\end{figure}

Figs.~\ref{fig:one} to ~\ref{fig:five} show the relative
astrometrically aligned VERA maps of $v$=1 and $v$=2
$J$=1$\rightarrow$0 SiO maser emission towards R\,LMi at the 4 epochs
of observations (see Table 1), except for epoch I in which only $v$=1
emission was observed.  We present first a brief description of the
maps followed by a discussion on the astrophysical implications. All
the figures are made from the data cubes, for which the thermal noises
are around 0.2\,Jy/beam per channel in both the $v$=2 and
$v$=1 datasets. We blanked all pixels less than 1 Jy/beam, and
averaged over the regions in which there was significant emission.

Fig.~\ref{fig:one} shows the map of $v$=1 maser emission at epoch I;
the structure in the map is very simple, with two groups of features
that we will refer to as the Northern and the Western clusters. The
$v$=2 maser observations failed due to a technical problem. 

Fig.~\ref{fig:two} shows the relative astrometrically aligned maps of
the $v$=2 and $v$=1 maser emission at epoch II, with respect to the
strongest channel in $v$=2. The compact features, with a velocity of
10.4\,\kms , seen at both transitions do overlap.

Fig.~\ref{fig:three} shows the maps at epoch III, the most feature
rich emission in both $v$=1 and 2, showing clusters to the North,
East, West and (in $v$=1 only) the South. We have used these
spots to draw a circle which represents the assumed ring of emission
around the central star. This ring is not a fitted solution, and it
serves as a useful guide to the eye for comparison between the spots
distribution at different epochs (epochs III and IV). 
The three clusters seen in both transitions (in Fig.~\ref{fig:three})
all show significant spatial offsets between spots with common
velocities.  We recall that coincidence of spots from different
transitions requires not only a negligible spatial offset, but also
coincidence in LSR velocity.  Both spatial offsets and velocities are
shown in Fig.~\ref{fig:three}, with arrows that indicate the distance
and direction between coincident spots at $v$=1 and $v$=2, and labels
for the corresponding velocities. The arrows head size corresponds to
the estimated relative position errors in our astrometric analysis.
The spots of the four regions with common emission at $v$=1 and 2 (two
in the North and one each in the Western and Eastern clusters) do not
overlap. The size of the spatial offsets between both transitions is
of several mas, except for the feature at -9.5 \kms, with an offset of
$0.5\,$mas, not much more than our estimated astrometric
errors. However, as Fig.~\ref{fig:last} (also epoch III) shows, this
region has a complex emission structure, with possible overlapping
regions very much outnumbered by those definitely not overlapping.

A map at epoch IV is shown in Fig.~\ref{fig:five}.  It shows some
emission in the Northern and Western regions. The observations are
only 27 days apart from epoch III and, albeit with far fewer spots,
the structure in epoch IV resembles the one in the map at epoch III.
We have superimposed the same circle as appears in
Fig.~\ref{fig:three} for visual comparison between the maps.

In summary, we have identified a total of about 17 spots in the $v$=1 line
and 21 spots in $v$=2 masers, from an inspection of the maps at epochs 
II,III and IV (where both $v$=1 and 2 masers were observed).
Among them we find two spots which are coincident within the conservative
astrometric estimates of uncertainties in our analysis, of 0.4 mas 
(the northern spot in epoch III, at -9.5 \kms\ LSR, 
and the dominant spot in epoch II). 
We  note that, since the displacements between nearby spots of different
transitions show in general very different directions and absolute
values, an error in the alignment of the maps could not explain the
poor spot coincidence. 

Our astrometric results are consistent with previous approachs
(Desmurs \etal\ 2000, Boboltz \& Wittkowski 2005, Yi \etal\ 2005, Cotton
\etal 2004, 2006, 2008), in that the radius of the $v$=1 is greater than the
$v$=2. These papers do not discuss the $v$=1,2 $J$=1$\rightarrow$0
spot coincidences and limit themselves to the average ring radii, as
they were based on less well founded map alignment procedures. Both
maser emissions occupy roughly the same regions, but the respective
spots are rarely coincident; in the observations presented in this
paper only about 1/10 of the identified spots are coincident to within
our error estimates (0.4\,mas).

The percentage of spot coincidence had been used as an argument favoring 
radiative or collisional pumping mechanisms. Although existing models really 
do not address the probability of intense common spots for these maser lines,
the lack of systematic coincidences tends to favor radiative pumping
because the conditions required to pump both masers are more different
in the radiative case than for collisional excitation 
(see details below and in the extensive discussion by Desmurs et al.\ 2000).

In the standard versions of both pumping schemes (see e.g.\ Bujarrabal
\& Nguyen-Q-Rieu, 1981; Lockett \& Elitzur, 1992), the inversion of
the $v$=1 and $v$=2 rotational transitions appear under quite
different conditions. This is due to the different excitation required
in both cases: the $v$=1 state is at 1800 K over the ground, while the
$v$=2 one is at 3600 K. Therefore, different (always high) excitation
would be required and both masers can rarely coexist in both
models. In the radiative model, the coincidence is somewhat less
probable than for the collisional pumping.  For instance, Bujarrabal
\& Nguyen-Q-Rieu estimate that in the radiative model the $v$=2 maser
pumping needs a density about 5 times larger than for the $v$=1, while
in the collisional scheme the factor is $\sim$2.2. This has led to
a long-lasting discussion on the consequences on theoretical
modelling of the possible systematic coincidence of $v$=1 and $v$=2
$J$=1$\rightarrow$0 spots. Both pumping mechanisms predict that masers
of rotational transitions within the same vibrational state
($J$=1$\rightarrow$0, $J$=2$\rightarrow$1, etc) must be strongly
coupled and should appear in practically the same points.
More recent theoretical studies address the general relative
distributions of the different maser spots, i.e. the ring radii for
the various lines. Humphreys et al. (2000, 2002) explains the slightly
larger radius found for the $v$=1 $J$=1$\rightarrow$0 maser assuming
collisional pumping driven by shocks; in their calculations the $v$=1
$J$=2$\rightarrow$1 ring is practically identical to the $v$=1
$J$=1$\rightarrow$0 one, as expected. Soria-Ruiz et al. (2004) can
also explain the difference observed between the two
$J$=1$\rightarrow$0 lines under the radiative excitation framework; in
this model, the position of the $v$=1 $J$=2$\rightarrow$1, 86-GHz maser
may vary depending on some possible alteration of the radiative
pumping (see below).

Our understanding of these problems has greatly changed from the first
empirical comparisons between $v$=1 $J$=1$\rightarrow$0 and $J$=2$\rightarrow$1
maser distributions (Soria-Ruiz \etal\ 2004, 2005, 2007). The $v$=1
$J$=2$\rightarrow$1 maser spots also occupy a ring, but are
systematically placed at a much larger distance from the star than
those of $v$=1 $J$=1$\rightarrow$0, and both spot distributions are
completely different.


Soria-Ruiz \etal\ (2004) explained these results invoking the effects
of line overlap between the ($v$=1,$J$=0) -- ($v$=2,$J$=1)
rovibrational transitions of SiO and the ($v_2$=0,12$_{7,5}$) --
($v_2$=1,11$_{6,6}$) rovibrational line of H$_2$O (see also Olofsson
\etal\ 1981, Bujarrabal \etal\ 1996). Their calculations indicate
that the overlap of these infrared lines reinforces the $v$=1,2
$J$=1$\rightarrow$0 masers; the quenching of the (otherwise intense)
$v$=2 $J$=2$\rightarrow$1 maser is also predicted, in agreement with
observations. The overlap also introduces a strong coupling between
$v$=1,2 $J$=1$\rightarrow$0 masers, that now must appear under very
similar conditions. Meanwhile, the $v$=1 $J$=2$\rightarrow$1 maser is
expected to be less affected by this phenomenon and to show
practically no relation with the two 43-GHz lines. The
calculations by Soria-Ruiz \etal\ (2004) are in agreement, at least
qualitatively, with all existing data on the relative large-scale distributions
 of the different SiO maser lines. In our opinion, at the moment it is not 
possible to understand the general properties of the maps of SiO maser 
maps without taking into account those theoretical considerations on 
the effects of line overlap.

We have seen that the results of radiative pumping models including
line overlap, as proposed by Soria-Ruiz \etal\ (2004), are compatible
with these (and other) observational results on SiO maser
distribution. But these models serve to understand the general
distribution of the SiO maser brightness, not to predict the
coincidence probability of the very intense and compact spots detected
in VLBI experiments. For such a purpose, we must understand the
competition between different maser lines under conditions of extreme
maser saturation, an intricate phenomenon that will be very difficult
to quantify. Therefore, we still lack for specific theoretical studies
on the probability of detecting, in both $v$=1 and $v$=2
$J$=1$\rightarrow$0, spots with very high brightness, which is the
true parameter to be compared with the spot coincidence ratio we can
deduce from the observations.  In summary, the very last observations,
made in response to the questions posed by the different theoretical
models, are now leading the modelling, and further efforts are
required to address the observational results.
In these efforts the capabilities of VERA, both for relative and
absolute astrometry, will facilitate major advances in the
observations of SiO masers.

\clearpage

\clearpage

\bigskip
\noindent
{\bf Acknowledgements}

\noindent
We would like to thank the VERA team for their support with the
observations.  We acknowledge support from the \emph{Spanish Ministry
of Education \& Science}, under grant PCI2005-A7-0246.
Richard Dodson acknowledges support as a Marie-Curie fellow via EU FP6
grant MIF1-CT-2005-021873.
We are grateful to Riccardo Cesaroni for providing single dish
measurements of water maser emission in several sources, which were 
useful to prepare the observations.  We acknowledge with thanks the 
variable star observations from the AAVSO
International Database, contributed by observers worldwide and used in
Figure 1. We would like to thank the anonymous referee for the comments
to this paper.


\end{document}